\documentclass[12pt]{article}
\usepackage{epsf,epsfig,amsmath}
\begin{document}

\title{Comment on Mermin's Recent Proof of the Theorem of Bell}

\author{Karl Hess$^1$ and Walter Philipp$^2$}

\date{$^1$ Beckman Institute, Department of Electrical Engineering
and Department of Physics,University of Illinois, Urbana, Il 61801
\\ $^{2}$ Beckman Institute, Department of Statistics and Department of
Mathematics, University of Illinois,Urbana, Il 61801 \\ }
\maketitle

\begin{abstract}

Mermin states in a recent paper that his nontechnical version of
Bell's theorem stands and is not invalidated by time and setting
dependent instrument parameters as claimed in one of our previous
papers. We identify a number of misinterpretations (of our
definitions) and mathematical inconsistencies in Mermin's paper
and show that Mermin's conclusions are therefore not valid: his
proof does not go forward if certain possible time dependencies
are taken into account.

\end{abstract}

We have presented a critique \cite{hp} of a nontechnical version
of Bell's theorem presented by Mermin \cite{mermin}. This critique
was based on the introduction of hidden time and setting dependent
instrument parameters in addition to the parameters considered by
Bell and others \cite{bellbook}. Mermin has responded to this
critique \cite{recmer} and has attempted to show that our time and
setting dependent instrument parameters fail to undermine his
reasoning.

We demonstrate below that Mermin \cite{recmer} has not properly
considered the stochastic independence of the family of our
instrument parameters from the source parameter. Owing to the
complexity of the problem, we will revert to notation similar to
that defined in our previous publications \cite{hp},
\cite{hpp1}-\cite{hpp3}. However, to facilitate the discussion, we
provide a one to one correspondence of Mermin's notation and ours,
at least as far as possible. We consider random variables $A=\pm1$
in station $S_1$ and $B=\pm1$ in station $S_2$ that describe the
potential outcome of spin measurements and are indexed by
instrument settings that are characterized by three-dimensional
unit vectors ${\bf a}, {\bf b}, {\bf c}$ in both stations. Mermin
introduces no precise counterpart for the random variables $A, B$
which describe potential outcomes and only considers actual
outcomes in form of green and red detector flashes. These green
and red flashes correspond then to values that $A, B$ assume: $+1$
which we can identify with green and $-1$ which we can identify
with red. The key assumption of Mermin \cite{recmer} and of Bell
and others \cite{bellbook} is that the random variables $A, B$
depend only on the setting in the respective station and on
another random variable $\Lambda$ that characterizes the spin of
particles emitted from a common source. Mermin uses instead of
$\Lambda$ instruction sets e.g. GGR meaning flash green for
settings ${\bf a}, {\bf b}$ (which Mermin actually labels $1, 2$)
and flash red for setting ${\bf c}$ (labelled $3$ by Mermin). In
our notation this means that for a certain parameter $\Lambda^*$
that corresponds to the instruction set GGR we have $A({\bf a},
\Lambda^*) = A({\bf b}, \Lambda^*) = +1$ and $A({\bf c},
\Lambda^*) = -1$. Please note that Mermin's notation does not
permit to distinguish (as is important in probability theory and
mathematical statistics) between the random variables, the values
these variables can assume and the experimental data. This
presents mathematical dangers as outlined by us previously
\cite{gref}. However, the discussion below can be completed
without going into this detailed distinction. The possible choices
of $\Lambda$ are restricted by Mermin \cite{recmer} and Bell
\cite{bellbook} invoking Einstein locality: $\Lambda$, or Mermin's
instruction sets, as well as their frequency of occurrence are
independent of the settings.

In addition to Mermin's (and Bell's) standard parameter $\Lambda$,
we have introduced setting and time $(t)$ dependent instrument
parameters which we denote here by ${\lambda}_{{\bf a},t}^*$,
${\lambda}_{{\bf b},t}^*$, ${\lambda}_{{\bf c},t}^*$ for station
$S_1$ and by ${\lambda}_{{\bf a},t}^{**}$, ${\lambda}_{{\bf
b},t}^{**}$, ${\lambda}_{{\bf c},t}^{**}$ for station $S_2$. These
parameters could be interpreted, for example, as computer programs
of two independent computers in the two stations. The computers
only need to be synchronized in time. We have discussed other
possible interpretations of these parameters in more detail in
\cite{hp} and \cite{hpp3}. Mermin named our instrument parameters
``microsettings" and we use this name for clarity below also.
However, our parameters are not microscopic in any sense. Their
main property is that they are time and setting dependent and are
stochastically independent of $\Lambda$. The union of Mermin's
parameters and our setting and time dependent parameters is termed
by Mermin \cite{recmer} the ``expanded instruction set". Note that
for any given setting and time of measurement there is exactly one
instrument parameter random variable (that can even represent the
result of a time dependent stochastic process) in each of the
stations $S_1$ and $S_2$ as defined above. Furthermore, these
additional parameters \cite{hp}, \cite{hpp1}-\cite{hpp3} are not
originating from the source but are specific to the instruments.
They may therefore have a frequency of occurrence that is
different from that of the source parameters and they may depend
on the setting in their respective station. Note that Mermin
remarks close to the end of his paper: ``The only complication
introduced by the microsettings of Hess and Philipp is that the
data actually produced ... is only determined when the particles
arrive in the neighborhood of their detectors and learn the
particular character of the conditions prevailing at the
detectors." and follows this statement by his footnote 12: ``Even
this minor complication is not really necessary". It will become
clear from the discussion below, that it is just the introduction
of setting and time dependent instrument parameters that extends
the parameter space beyond Bell's and makes all the difference. By
instruments we mean the whole system of polarizers and detectors
that is necessary to perform the measurements. These instrument
parameters invalidate Mermin's reasoning. Of course, if only
source parameters are considered and no time and setting dependent
instrument parameters are included both Bell's and Mermin's proofs
stand even if certain time dependencies of $\Lambda$ are included.
Obviously this is not the case, in general, as can be seen from
the simple example: $\Lambda \equiv t$ and $A,B$ depend on time
$t$. It is important to realize in connection with footnote 12 of
Mermin that time is not just an information that is sent out from
the source and therefore can be included into the parameter
$\Lambda$. There may be time dependent processes in the
instruments that give rise to frequencies of occurrence of
instrument parameters that have nothing to do with $\Lambda$ and
that can depend on the setting. We interpret some of the writings
of Mermin (and of others) as follows: To include all the effects
of time $t$ in the source parameter, the original $\Lambda$ may be
replaced by the pair $(\Lambda, t)$. This pair is sent out to the
stations $S_1$ and $S_2$. It is then understood that the pair
interacts with the settings (e.g. ${\bf a}, {\bf b}$ in stations
$S_1$ and $S_2$, respectively) thereby determining the station
parameters ${\lambda}_{{\bf a},t}^*$ and ${\lambda}_{{\bf
b},t}^{**}$ as $\bf {functions}$ of $(\Lambda, t, {\bf a})$ and of
$(\Lambda, t, {\bf b})$, respectively. This obviously contradicts
the fact that ${\lambda}_{{\bf a},t}^*$ and ${\lambda}_{{\bf
b},t}^{**}$ are stochastically independent of $\Lambda$.

As far as the nature our instrument parameters is concerned we add
the following explanation. They could be visualized as computer
programs in the measurement stations. We maintain that proofs of
the Bell theorem that cannot accommodate the possibility of such
computer programs have a much reduced importance.

Mermin also adds another footnote (13) in which he claims that
``detections separated by arbitrary long time intervals" force the
instruction sets back to his of reference \cite{mermin}. We will
return to this point below but remark here that our model is
highly flexible with respect to separations of the detection time
of the correlated pair and can accommodate such separations with
ease. We also remark that no experiments exist that encompass
arbitrary long separations of detection times.

Consider also the following additional elements of our theory.
Assume we have settings ${\bf a}, {\bf a}$ in $S_1$ and $S_2$,
respectively, corresponding to Mermin's settings $11$, at the
times $t_{i1}, i = 1,2,...L$ when $L$ such measurements are made.
Also assume that we have settings ${\bf a}, {\bf b}$ corresponding
to Mermin's $12$ at the times $t_{j2}, j = 1,2,...L$. There is no
loss of generality to assume that the number of occurrences of the
settings $11$ and $12$ is the same as an easy application of the
strong law of large numbers shows. Obviously, all these $2L$ time
points of the $2L$ measurements must be different if one wants to
consider measurements as described in \cite{gwzz}, \cite{eprex}.
Thus, if the station parameters depend on time, all the $2L$
random variables, indexed by $t_{i1}$ and $t_{j2}$ could be
different. Consequently all the random variables $A$ and $B$
corresponding to these time points could be different. As an
example we could have:
\begin{equation}
A({\lambda}_{{\bf a},{t_{i1}}}^*, \Lambda) \neq A({\lambda}_{{\bf
a},{t_{j2}}}^*, \Lambda) \text{ for } i,j = 1,2,...L \label{mf1}
\end{equation}
while we still have that with probability one
\begin{equation}
A({\lambda}_{{\bf a},{t_{i1}}}^*, \Lambda) = B({\lambda}_{{\bf
a},{t_{i1}}}^{**}, \Lambda) \text{ for } i = 1,2,...L \label{mf2}
\end{equation}
where $A = B$ means the same color in both stations.

The ${\lambda}_{{\bf a},{t_{i1}}}^*, i = 1,2,...L$ may be
interpreted as corresponding to Mermin's term microsettings.
However, our expanded instruction sets do not collapse back onto
his instruction sets. We emphasize again, that the family of our
instrument parameters is stochastically independent of the source
parameter $\Lambda$. Thus, in this context, the crucial task is to
properly choose the $\bf {stochastically}$ independent variables
for the given physical problem.

We believe we have shown, here and in all of our previous
publications, that it is not possible to exclude a time dependence
of the parameters describing EPR experiments \cite{gwzz},
\cite{eprex} on the basis of the statistical data of the given
problem. We also are not aware of any physics that excludes time
dependencies e.g. in the physics of gyroscopes on a rotating earth
\cite{hpp3} or the electrodynamics of moving bodies in general.
One thus is faced with the following fact: the experiments may
contain time dependencies.  Bell's and Mermin's model disregard
general time dependencies, particularly stochastically independent
(of $\Lambda$) instrument parameters that depend on both time and
setting. The model of Gill et. al. \cite{gwzz} which is also
alluded to by Mermin \cite{recmer} declares time as irrelevant.
However, this is a postulate not a proof. Because the experiments
may contain time dependencies, we can only conclude that these
models may be irrelevant.

Up to now we have assumed detection times of the correlated pair
that are approximately equal i.e. occur within the same short time
interval in the laboratory reference frame. Our discussion can, of
course, also accommodate detections separated in time in contrast
to the claim of Mermin in his footnote 13 of reference
\cite{recmer}. Just choose the zero of the time scale in each
station at the time (or within the short time interval) of
detection of the first correlated pair and everything remains
unchanged. We only add that experiments with detections separated
by considerable time periods (say minutes or hours) have not been
performed.

For completeness we add the definition of Einstein locality and
the following remarks concerning the differences between our model
and that of Bell (of which Mermin's model is just a special case).

\begin{itemize}

\item[(a)] We define Einstein locality by the following postulate:
no influence can be exerted by actions in one station $S_1$ on
events in a spatially separated station $S_2$ (and vice versa)
with a speed that exceeds that of light in vacuo.

\item[(b)] We assume that the experiments correspond to the ideal
assumptions of Bell: The actual instrument settings ${\bf a}, {\bf
b}$ and ${\bf c}$ are chosen randomly (the randomness being
guaranteed according to taste by a computer, a person with free
will, a quantum mechanical measurement system or Tyche) and after
the correlated pair has been emitted from the source (delayed
choice \cite{eprex}).

\item[(c)] The conditions in Bell's mathematical model are not
necessary to fulfill these definitions and assumptions. Bell's
conditions, however, are sufficient and can be expressed as
follows: the source parameters $\Lambda$ do not depend on the
settings ${\bf a}, {\bf b}, {\bf c}$ and the functions $A, B$ only
depend on the actual setting of their respective station and not
on that of the other. Bell further requires that $\Lambda$ has a
probability distribution $\rho$ that remains unchanged over the
whole run of experiments. For a given setting in each station,
this means that the random variables $A$ and $B$ occur with
frequencies that are related to $\rho$ in both stations. It is
instructive to regard the pair $\Lambda, {\bf a}$ on which the
function $A$ depends as a new setting dependent parameter. Then
Bell's approach does contain correlated setting dependent
parameters. However, their density is identical to that of
$\Lambda$. Following work of Jarret et.al. \cite{nus}, Bell
\cite{bellbook} also included additional but completely
uncorrelated station-specific parameters. In contrast, our station
parameters have a density that is different from that of $\Lambda$
and they are correlated through time-dependencies. We have shown
repeatedly that Bell type proofs do not go through when such
parameters are involved \cite{hp}-\cite{hpp1}, \cite{hpp3}.

\item[(d)] Our mathematical model is therefore more general than
that of Bell and also obeys Einstein locality. We have added
setting and time-dependent parameters ${\lambda}_{{\bf a},t}^*$,
${\lambda}_{{\bf b},t}^*$, ${\lambda}_{{\bf c},t}^*$ for station
$S_1$ and ${\lambda}_{{\bf a},t}^{**}$, ${\lambda}_{{\bf
b},t}^{**}$, ${\lambda}_{{\bf c},t}^{**}$ for station $S_2$ that
can, in any sequence of experiments, have a frequency of
occurrence or density that is independent of that of $\Lambda$.
This density will depend only on the setting of the respective
station and not on the density of the source parameters. The
time-dependence of the parameters suffices also to fulfill certain
additional requirements e.g. that we have for the same setting
$\bf a$ and time $t$: $A({\bf a},t) = B({\bf a},t)$ (recall that
$A = B$ means same colors). Clearly, our mathematical model
contains that of Bell as a special case and still obeys Einstein
locality. Naturally clocks in the different stations can show
correlated, even identical, times without violating Einstein
locality which only requires that no influences are exerted with a
speed faster than that of light in vacuo.

\end{itemize}

Thus, our model goes beyond that of Bell and fulfills Einstein
locality \cite{foot1}. Mermin's proof, on the other hand, does not
go forward if time and setting dependent station parameters are
included.

Acknowledgement: the work was supported by the Office of Naval
Research N00014-98-1-0604 and the MURI supported by ONR.


\begin{thebibliography}{99}
\bibitem{footc1} Of course, the functions $A$ on the left and on
the right side of Eq.(\ref{me2}) are not the same in the strict
mathematical sense. We hope that this abuse of notation will not
cause confusion.
\end{thebibliography}
\end{document}